\documentclass[a4paper,11pt]{article}

\usepackage{jheppub}

\usepackage[utf8]{inputenc}
\usepackage{amsmath}
\usepackage{graphicx}
\usepackage{color}
\usepackage[dvipsnames]{xcolor}
\usepackage[version=3]{mhchem}

\definecolor{nicegreen}{rgb}{0.1,0.5,0.1}
\definecolor{darkblue}{rgb}{0.15, 0.2, .85}
\definecolor{darkgreen}{rgb}{0.1,0,0.3}
\definecolor{darkred}{rgb}{0.6,0,0}
 
\newcommand{\sinsq}[1]{\sin^2 \theta_{#1}} 
\newcommand{\Deltamsq}[1]{\Delta m^2_{#1}}
\newcommand{\eV}{\, \rm eV}

\newcommand{\mdm}{m_\chi}

\newcommand{\Eth}{E_{\rm th}}

\newcommand{\MeV}{\, {\rm MeV}}
\newcommand{\GeV}{\, {\rm GeV}}

\newcommand{\cm}{\, {\rm cm}}

\newcommand{\m}{\, {\rm m}}
\newcommand{\g}{\, {\rm g}}

\newcommand{\RID}{R_{\textrm{ID}}}
\newcommand{\hID}{h_{\textrm{ID}}}

\newcommand{\GENIE}{{\tt GENIE }}
\newcommand{\nuCraft}{{\tt nuCraft }}
\newcommand{\ROOT}{{\tt ROOT }}
\newcommand{\wimpsim}{{\tt WimpSim }}

\newcommand{\Rsun}{R_\odot}
\newcommand{\Dsun}{D_\odot}

\newcommand{\sigmaSD}{\sigma_{\chi p}^{\rm SD}}

\definecolor{ShamrockGreen}{rgb}{0.0, 0.62, 0.38}

\title{Searching for Dark Matter in the Sun using Hyper-Kamiokande}

\author[a]{Nicole F. Bell,}
\author[a]{Matthew J. Dolan,}
\author[a]{Sandra Robles}
\affiliation[a]{ARC Centre of Excellence for Dark Matter Particle Physics, \\
School of Physics, The University of Melbourne, Victoria 3010, Australia}
\emailAdd{n.bell@unimelb.edu.au}
\emailAdd{matthew.dolan@unimelb.edu.au}
\emailAdd{sandra.robles@unimelb.edu.au}

\abstract{
We study the ability of the Hyper-Kamiokande (HyperK) experiment, currently under construction, to constrain a neutrino signal produced via the annihilation of dark matter captured in the Sun. We simulate upward stopping and upward through-going muon events at HyperK, 
using Super-Kamiokande (SuperK) atmospheric neutrino results for validation, together with fully and partially contained events. Considering the annihilation of dark matter to various standard model final states, we determined the HyperK sensitivity to the dark matter spin-dependent scattering cross-section. We find that HyperK will improve upon current SuperK limits by a factor of 2-3, with a further improvement in sensitivity possible if systematic errors can be decreased relative to SuperK. 
}

\begin{document}
\maketitle

\section{Introduction}
\label{sec:intro}

The nature of particle dark matter (DM) continues to elude us. One paradigm for identifying the properties of dark matter is via indirect detection, where we search for DM annihilation or decay products in regions of high dark matter density, such as the Galactic centre or dwarf spheroidal galaxies.
Another possibility is that dark matter can be captured by stars. This includes capture in the Sun~\cite{Press:1985ug,Silk:1985ax,Srednicki:1986vj,Gould:1987,Griest:1986yu,Krauss:1985ks,Gould:1987ju,Jungman:1995df,Peter:2009mk}, as the Solar System travels through  the Galactic dark matter halo.
This can lead to observable signals if the accumulated DM annihilates either to Standard Model (SM) final states that have neutrinos in their decay chains~\citep{Super-Kamiokande:2004pou,Tanaka:2011uf,Choi:2015ara,Adrian-Martinez:2016ujo,Adrian-Martinez:2016gti,Aartsen:2016zhm,Boliev:2013ai}, or to dark sector particles that escape the Sun~\cite{Batell:2009zp,Schuster:2009au,Bell:2011sn, Feng:2016ijc,Leane:2017vag,Albert:2018jwh,Bell:2021pyy}.

The annihilation of solar DM to neutrinos can be used to constrain both the spin-independent and spin-dependent DM-nucleon scattering cross-sections, $\sigma_{\chi p}^{\rm SI}$ and $\sigmaSD$, respectively. While the limits that can be set on the spin-independent interactions are much weaker than those arising from direct detection experiments such as XENON1T~\cite{Aprile:2018dbl,Aprile:2020thb} and PandaX-4T~\cite{PandaX:2021osp}, the solar-capture technique can set competitive limits on spin-dependent interactions, for certain DM masses.

Searches for neutrinos arising from DM annihilation in the Sun are well established. The IceCube~\cite{Aartsen:2016exj,Aartsen:2016zhm} and Super-Kamiokande (SuperK)~\cite{Choi:2015ara} experiments currently place the leading bounds on $\sigmaSD$, at higher and lower DM masses respectively, and there are also existing limits from Antares~\cite{Adrian-Martinez:2016gti} and BAKSAN~\cite{Boliev:2013ai}. In this article we study the ability of the forthcoming Hyper-Kamiokande (HyperK) neutrino experiment to constrain dark matter annihilation in the Sun, focusing on the spin-dependent scattering cross-section $\sigmaSD$. HyperK will be a large water Cherenkov detector located in the Tochibora mine near Kamioka in Japan, and will be the successor to the highly successful SuperK experiment. It is due to commence physics operations in 2027. It is therefore timely to consider HyperK's ability to constrain the properties of dark matter.

In previous work, we explored the constraints that HyperK can set on the thermally averaged annihilation cross section of sub-GeV dark matter which annihilates in the Galactic centre~\cite{Bell:2020rkw}, and demonstrated that HyperK will be able to probe the annihilation of light DM to neutrinos down to thermal-relic scale cross-sections. The HyperK Design Report~\cite{Abe:2018uyc} includes sensitivity projections for DM annihilation to neutrinos in the Galactic Centre, and for the annihilation of DM accumulated in the Earth.  However, the Design Report does not consider the annihilation of DM accumulated in the Sun, the topic of this work.

In ref.~\cite{Bell:2020rkw}, we used the fully contained and partially contained event categories in simulations of SuperK and HyperK.  In the present paper, we also include the contributions from upward-going muons which can either stop within or pass through the inner detector. These upward-going muons are the most important event categories for dark matter annihilation in the Sun. Similar studies for SuperK can be found in refs.~\cite{Kappl:2011kz,Baum:2016oow}.

We focus on heavier Standard Model final states whose decay chains include neutrinos: $b\bar{b}$, $W^+W^-$ and $\tau^+\tau^-$. We also study annihilation directly to neutrinos, $\chi\chi \to \nu\bar{\nu}$. Other SM final states, such as muons and lighter quarks, lose energy and decay at rest (light quarks first hadronise to states including pions, whose decays involve neutrinos), which leads to neutrinos with energies between 10 and 100~MeV. Therefore, limits on dark matter annihilation to those final states can be obtained via searches for these energy neutrinos~\cite{Bernal:2012qh,Rott:2012qb}. Light dark matter, with mass less than a few GeV, can also lead to heat transfer within the Sun leading to noticeable effects on solar physics~\cite{Spergel:1984re,Gould:1989hm,Vincent:2013lua,Busoni:2017mhe}. Dark matter which is lighter than 4~GeV is strongly affected by evaporation in the Sun \cite{Griest:1986yu,Gould:1987,Hooper:2008cf,Busoni:2013kaa,Garani:2017jcj,Busoni:2017mhe}, and so we do not consider dark matter with mass less than 4~GeV. Hence we do not consider the impact of heat transfer.

We next provide some theoretical background on dark matter capture and annihilation in the Sun, before discussing details of our simulations (section~\ref{sec:sim}) and our projections for the constraints that HyperK can set, compared with other experiments (section~\ref{sec:results}).

\section{Dark Matter Capture in the Sun}
\label{sec:dm}

To calculate the signal neutrino flux, we first estimate the DM annihilation rate.
The neutrino flux from DM annihilation in the Sun at the detector location is given by
\begin{equation}
\dfrac{d\Phi_\nu}{dE_\nu} = \frac{\Gamma_A}{4\pi\Dsun^2 }    \dfrac{dN_\nu}{dE_\nu} \, ,
\label{eq:neutrino_flux}
\end{equation}
where $\Gamma_A$ is the annihilation rate, $\Dsun$ is the distance from the  detector to the Sun and $dN_\nu/dE_\nu$ 
is the neutrino spectrum per dark matter annihilation.

The time evolution of the total number of DM particles, $N_\chi$, in the Sun is determined by
\begin{equation}
\dfrac{dN_\chi}{dt} = C - E N_\chi - A N_\chi^2 ,
\label{eq:NWIMPs}
\end{equation}
where $C$ is the capture rate, $A$ is related to the annihilation rate via $\Gamma_A=\frac{1}{2}AN_\chi^2$, and $E$ is the evaporation rate. Evaporation is expected to have no impact for DM masses $\mdm\gtrsim 4\GeV$ \cite{Griest:1986yu,Gould:1987,Hooper:2008cf,Busoni:2013kaa,Garani:2017jcj,Busoni:2017mhe}. Consequently, we do not attempt to set limits on DM masses lower than 4~GeV.
Neglecting $E$, we can solve Eq.~\ref{eq:NWIMPs} to find
\begin{equation}
N_\chi= \sqrt{\frac{C}{A}} \tanh\left(\frac{t}{\tau}\right),    
\end{equation}
and hence 
\begin{equation}
\Gamma_A = \frac12 A N_\chi^2 = \frac12 C \tanh^2\left(\frac{t}{\tau}\right),    
\end{equation}
where
\begin{equation}
\tau = \frac{1}{\sqrt{CA}}.     
\end{equation}
For $t/\tau\gg1$, capture and annihilation are in equilibrium, such that the number of DM particles in the Sun remains constant, $dN_\chi/dt = 0$, and the annihilation rate is equal to half the capture rate,
\begin{equation}
\Gamma_A = \frac12 C.    
\end{equation}
The equilibrium time scale depends on both the DM annihilation and DM-nucleon scattering cross sections. If the annihilation cross section is of order that for a thermal relic DM candidate, capture-annihilation equilibrium is easily satisfied for all DM-nucleon cross-sections of interest in this work. This removes all dependence on the annihilation cross section, and allows limits to be expressed in term of the DM-nucleon scattering cross section alone, which can be directly compared with limits arising from direct detection experiments. As in the existing SuperK analyses~\cite{Tanaka:2011uf,Choi:2015ara}, we shall work in the  $t/\tau\gg1$ limit throughout.

The total rate at which DM is captured in the Sun is~\cite{Gould:1987ir}
\begin{equation}
C = \int_0^{\Rsun} 4\pi r^2 dr  \sum_i^{N_{\rm{spec}}}  \int_0^{u^{max}_\chi} du_\chi \frac{f(u_\chi)}{u_\chi} \,w \, \Omega_{i}^-(r), 
\end{equation}
where $\Omega_{i}^-(r)$ is the probability of being captured by scattering on the nuclear species $i$, $N_{\rm{spec}}$ is the number of nuclear species in the Sun,  $u_\chi$ is the DM velocity far away from the Sun,  $f(u_\chi)$ is the DM velocity distribution, $v_{esc}$ is the escape velocity at the interaction radius and $w(r)=\sqrt{u_\chi^2+v_{esc}(r)^2}$ is the DM speed at the same radius. In principle $u^{max}_\chi=\infty$. 
The capture probability depends on the scattering cross section, including form factors for the different nuclei, the kinematics of the scattering process and the nuclear species abundance~\cite{Busoni:2017mhe},
\begin{equation}
 \Omega_{i}^-(r) =  \sigma_{\chi i} \, n_i(r) \frac{4\mu_+^2}{\mu w}   \int_{\frac{w |\mu_-|}{\mu_+}}^{v_{esc}} dv\, v\, |F_i(q_{tr})|^2, 
\end{equation}
where 
\begin{equation}
\mu=\frac{\mdm}{m_i}, \qquad \mu_\pm=\frac{\mu\pm1}{2}, 
\end{equation}
$F_i(q_{tr})$ is the nuclear form factor, $q_{tr}$ the momentum transfer, $m_i$ is the mass of the nuclear species $i$ and
\begin{equation}
\sigma_{\chi i} = \sigma_{\chi N} A_i^2 \left(\frac{m_i}{m_N}\right)^2 \left(\frac{\mdm + m_N}{\mdm+m_i}\right)^2,    
\end{equation}
where $A_i$ is the atomic number of the nucleus $i$, $m_N$ is the nucleon mass and $\sigma_{\chi N}$ is the DM-nucleon scattering cross section.

\section{Simulation and Validation}
\label{sec:sim}

 In this section we briefly summarise the detector simulation and validation undertaken in our previous work~\cite{Bell:2020rkw}; full details are contained therein.  We then focus on what is new to this work: the incorporation of the event categories including upward-going muons. 
 
 We simulate both the SuperK and HyperK detectors using the \ROOT geometry package~\cite{Brun:1997pa}, using material properties from \texttt{GEANT4}~\cite{Brun:1994aa}. We use the \GENIE3.0.4a~\cite{Andreopoulos:2009rq, Andreopoulos:2015wxa}  package with the G18\_10a tune to model the interactions of neutrinos with matter such as the detector, water  or surrounding rock. \GENIE calculates all the relevant cross-sections and neutrino production mechanisms such as quasi-elastic scattering, resonant production and deep inelastic scattering. The deep inelastic scattering cross sections are calculated with the GRV98 LO PDF~\cite{Gluck:1998xa}, using \texttt{LHAPDF5} \cite{Whalley:2005nh}.
 We use the  HKKM11~\cite{Honda:2011nf} atmospheric neutrino fluxes as our primary background.

 For the atmospheric neutrino fluxes we include the effect of neutrino oscillations in the Earth using the Preliminary Reference Earth Model~\cite{Dziewonski:1981xy} within the \nuCraft code~\cite{Wallraff:2014qka}. We take the neutrino oscillation parameters from the Particle Data Group (PDG)~\cite{Zyla:2020zbs}, as specified in Table~\ref{tab:oscparams} below, assuming the normal mass hierarchy. Since we do not consider dark matter masses lower than 4~GeV in this paper we do not consider the diffuse supernova or spallation backgrounds, or the impact of neutron tagging, all of which are relevant at very low energies. Our simulation and validation against SuperK data for the Fully Contained (FC) and Partially Contained (PC) event categories for electrons and muons can be found in ref.~\cite{Bell:2020rkw}.
 
\begin{table}[h]
\centering
\begin{tabular}{|c|c||c|c|}
\hline
Parameter & Value & Parameter & Value \\
\hline
$\sinsq{12}$ & $0.307 \pm0.013$ & $\Deltamsq{21}$ & $(7.53 \pm 0.18) \times 10^{-5} \eV^2$  \\
$\sinsq{23}$ & $0.545 \pm 0.021$  & $\Deltamsq{32}$ & $(2.453 \pm 0.034) \times 10^{-3} \eV^2$ \\
$\sinsq{13}$ & $0.0218 \pm 0.0007$ & $\delta$ & $(1.37 \pm {0.17})$ $\pi$~rad\\
\hline
\end{tabular}
\caption{Neutrino parameters from the PDG~\cite{Zyla:2020zbs}, assuming normal mass ordering. }
\label{tab:oscparams}
\end{table}

\subsection{Signal Yield}
The physics of DM capture and annihilation in the Sun has been extensively studied, as has the subsequent propagation of the DM annihilation products inside the Sun. There are a number of well-tested codes available which incorporate this physics, such as \texttt{WimpSim}~\cite{Blennow:2007tw,Edsjo:2017kjk,Niblaeus:2019gjk}, \texttt{DarkSUSY}~\cite{Gondolo:2004sc,Bringmann:2018lay}, and $\chi$\texttt{aro}$\nu$~\cite{Liu:2020ckq}.

We use \texttt{DarkSUSY}~v.6.2~\cite{Gondolo:2004sc,Bringmann:2018lay} to calculate the capture rate, performing a full numerical integration over the DM velocity distribution and the momentum transfer in the form factors. Depending on the annihilation channel considered, we study different DM mass ranges for our HyperK projections, as specified in Table~\ref{tab:massrange}. For comparison with SuperK we use the more limited range of masses given in Table~1 of ref.~\cite{Choi:2015ara}. For the DM masses in Table~\ref{tab:massrange}, we compute the capture rate for spin-dependent (SD) scattering off protons assuming $\sigmaSD=10^{-41}\cm^2$ and $\rho_\chi = 0.3 \GeV \cm^{-3}$ (as in the SuperK analysis~\cite{Choi:2015ara}).  

\begin{table}[t]
    \centering
    \begin{tabular}{|l|c|}
    \hline
       Channel & $\mdm$ (GeV)  \\
     \hline      
        $\chi\chi\rightarrow\nu \overline{\nu}$  & 4 - 500\\
        $\chi\chi\rightarrow\tau^+\tau^-$  & 4 - 500  \\  
        $\chi\chi\rightarrow b\overline{b}$  & 5.5 - 500 \\           
        $\chi\chi\rightarrow W^+W^-$ & 80.3 - 500  \\       
     \hline         
    \end{tabular}
    \caption{The dark matter mass ranges used for generating signal events per annihilation channel. }
    \label{tab:massrange}
\end{table}

The DM capture rate is subject to uncertainties in the solar composition and in the nuclear form factors. In the case of spin-dependent interactions, these uncertainties are expected to be small~\cite{Choi:2015ara}. However, for spin-independent couplings they can be substantial - up to 25\% for the solar model and 45\% for the form factors~\cite{Choi:2015ara}. The SuperK analysis~\cite{Choi:2015ara} uses the  BS2005-OP and BS2005-AGS,OP~\cite{Bahcall:2004pz} solar models.

We generate the neutrino yield, $dN_\nu/dE_\nu$, from DM annihilation in the Sun using the \wimpsim v5.0~\cite{Blennow:2007tw,Edsjo:2017kjk,Niblaeus:2019gjk} package.
The \texttt{WimpAnn} program within \wimpsim simulates WIMP annihilations in the Sun, taking into account several effects, such as hadronic interactions of the annihilation products in the centre of the Sun, and the interactions and oscillations of the neutrinos as they propagate through the Sun. \texttt{WimpAnn} uses the density profile of the Sun given in the AGSS2009 solar model~\cite{Serenelli:2009yc}.  We consider the $b\overline{b}$, $W^+W^-$, $\tau^+\tau^-$ and $\nu \bar{\nu}$ annihilation channels, for the dark matter masses in Table.~\ref{tab:massrange}, and generate $10^6$ events per mass and annihilation channel. 

The coordinates of the detector and a time-frame to generate events are provided as inputs of the simulation so that a time stamp is assigned to every event.
With this information, the Sun-Earth distance (relevant for oscillations in vacuum) is calculated using
the positional astronomy library \texttt{SLALIB}~\cite{SLALIB:2014}, and the amount of material in the Earth the neutrino has to pass through is determined. The
oscillation parameters are an input of \texttt{WimpAnn}, so there is no need to provide \texttt{WimpEvent} with these
inputs. To include neutrino oscillations in the Earth, \texttt{WimpEvent} uses the PREM model~\cite{Dziewonski:1981xy}. Once the events are projected on to flavour eigenstates at the detector location, the neutrino yield is calculated considering charged and neutral currents, with water as the target material. 
 The location of the HyperK detector is taken to be ($36^\circ\,21^\prime\,20.105"$N, $137^\circ\,18^\prime\,49.137"$E)~\cite{Abe:2018uyc}.

The background HKKM11 flux is binned in energy, $\cos z$ and azimuth. We bin the solar neutrino flux from DM annihilation in the same way, namely 20 bins per decade in energy,  20 bins in $\cos z$, and 12 bins in azimuth. The number of energy bins depends on the DM mass, and spans in the range  $E_\nu\in [10\MeV,10^{\log(\lceil\mdm\rceil}]$, where $\lceil\mdm\rceil$ is the ceiling function. 
The binned flux obtained at the detector location (corresponding to Eq.~\ref{eq:neutrino_flux}), normalised to a spin-dependent cross-section of $\sigmaSD=10^{-41}\cm^2$ is 
\begin{equation}
\dfrac{d\Phi_\nu^{i,j,k}}{dE_\nu} = \frac{1}{2} C_\odot(\sigmaSD=10^{-41}\cm^2) \, \frac{dN_\nu}{dE_\nu}  (E_\nu^i,\cos z^j,a^k)\Big|_{\texttt{WimpSim}}. 
\label{eq:DMfluxbinned}
\end{equation}
The factor of $1/4\pi\Dsun^2$ in Eq.~\ref{eq:neutrino_flux} is not explicitly shown in Eq.~\ref{eq:DMfluxbinned} above, as this factor is included in the result from \texttt{WimpEvent}, i.e., it is absorbed into the definition of $\frac{dN_\nu}{dE_\nu}\Big|_{\texttt{WimpSim}}$. We take the oscillated flux in Eq.~\ref{eq:DMfluxbinned}
as the input of our \texttt{GENIE}-based detector simulation  for FC and PC events. We describe our treatment of upward-going muons below.

\subsection{Upward-going Muons}

Along with the detector itself, neutrinos interact with the rock surrounding the detector. This means that muon neutrinos can produce highly energetic muons, via charged-current interactions, well outside the detector. Provided that the energy of the incoming neutrino is high enough, the muon produced by the neutrino will enter the detector and emit Cherenkov light. While muons traveling in the downward direction cannot be discriminated from the overwhelming cosmic-ray background, those moving upwards can be unambiguously identified as neutrino induced and are called upward-going muons.

Upward-going muons can be further classified as follows: Stopping upward-going muons come to rest in the detector, while through-going muons traverse the entire detector. Through-going muons can be showering or non-showering. Showering upward-going muons have accompanying radiation since radiative energy loss is, in this case, the leading muon energy loss mechanism. Non-showering events do not undergo radiative energy loss. Neutrinos with energies $\sim 10$ GeV produce stopping upward-going muons. This is a similar energy range to that of partially contained events. In contrast, through-going muons are significantly  more energetic, with an energy range around 100 GeV for non-showering muons, and more than 1 TeV for showering upward-going muons.

Upward-going muon events are produced via neutrino interactions in the surrounding rock as well as in the water of the outer detector. 
To simulate the muons which reach the detector, we assume that interactions take place in ``standard rock'' as defined by SuperK, with $Z=11$, $A =22$ and $\rho=2.65\g \cm^{-3}$. We restrict the neutrino interaction point to be within 4~km of the centre of the detector, which is sufficiently large given the muon range of the highest energy atmospheric neutrinos~\cite{Lipari:1991ut}.

Through-going muon events have both entrance and exit signals in the outer detector, while upward stopping events have only an entrance signal. In addition, SuperK requires through-going muons events to have a trajectory greater than 7~m within the inner detector, to reduce the background of photo-production of pions by energetic muons that pass nearby the detector~\cite{Ambrosio:1997qh}. The same path-length cut is imposed on the stopping events~\cite{Ashie:2005ik}. The minimum path length corresponds to a minimum energy, $\Eth$, for muons entering the detector. Muons with energy below this threshold will stop in a distance less than 7~m and hence not form part of our analysis. We assume the same path-length cuts for HyperK. We adopt an analytic approach to determine the through-going muon flux, with the stopping muon flux then taken as the difference between this and the total flux. 

\subsection{Through-going muons: Analytic method}

To estimate the upward-going muon flux, we follow the method outlined in ref.~\cite{Hatakeyama:1998fln}, which was adapted from the upward-going muon flux calculation for supernova SN1987A~\cite{Honda:1987ig}. At high energy, the average rate of muon energy loss  in a material can be written as~\cite{Barrett:1952woo}
\begin{equation}
    \left\langle - \dfrac{dE_\mu}{dX} \right\rangle = a(E_\mu) + b(E_\mu) E_\mu,
\end{equation}
where $E_\mu$ is the total energy, $a(E_\mu)$ is the ionization energy loss, and $b(E_\mu)$ is due to radiative processes, Bremsstrahlung, $e^+e^-$ pair production, and photonuclear interactions.
The mean stopping power for muons in several materials is listed in  Tables~II-28 (liquid water) and IV-6 (standard rock)\footnote{Tables available online at  \href{https://pdg.lbl.gov/2021/AtomicNuclearProperties/index.html}{pdg.lbl.gov/AtomicNuclearProperties/}.} of ref.~\cite{Groom:2001kq}.

\begin{figure}
    \centering
    \includegraphics[width=0.6\textwidth]{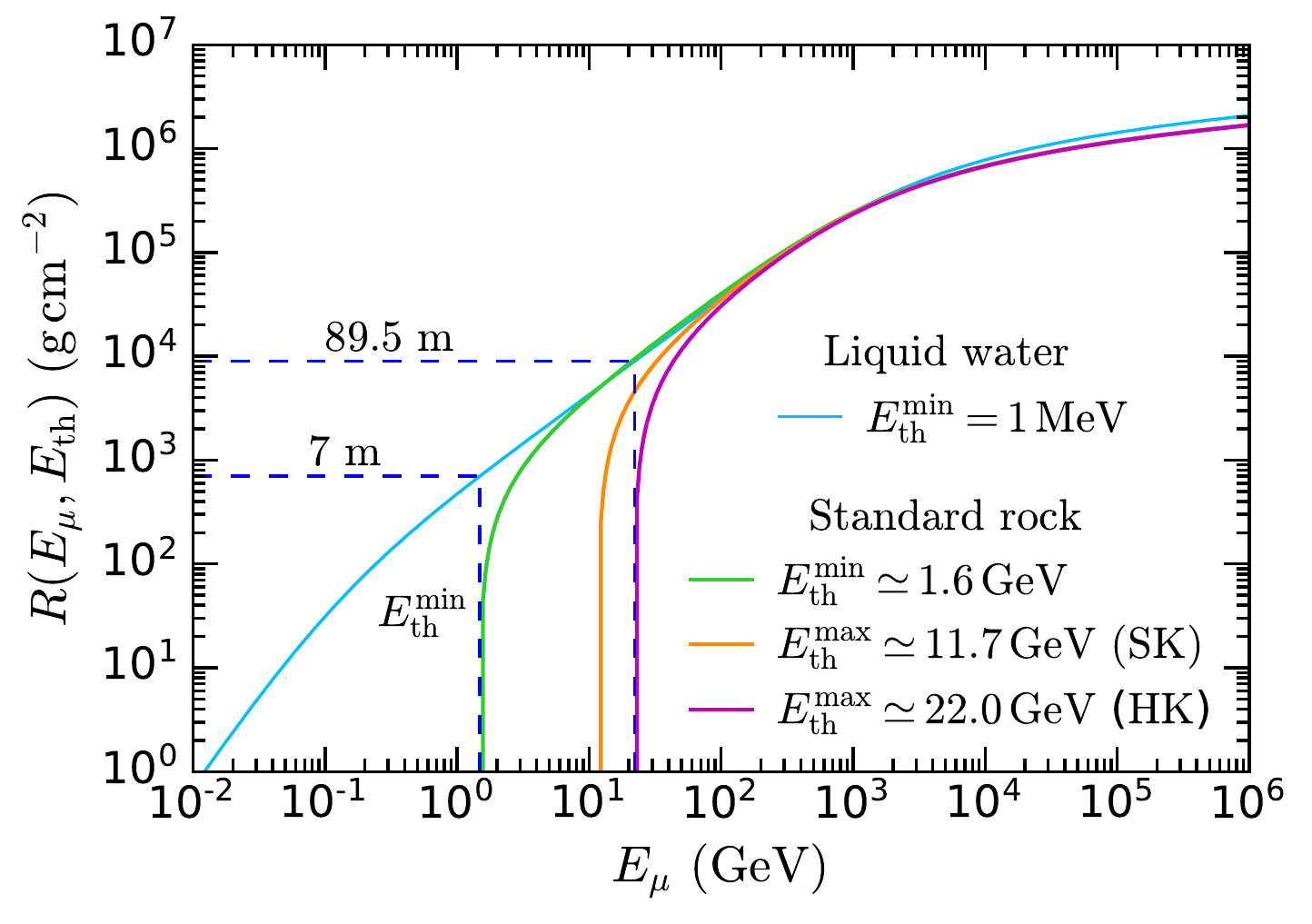}
    \caption{Muon range in liquid water (light blue) and in standard rock for the minimum ($7\m$, green, corresponding to $\Eth=1.6$~GeV) and maximum ($89.5\m$, magenta) track length in HyperK.
    The maximum muon path length in SuperK ID is shown in orange. }
    \label{fig:muonrange}
\end{figure}

We assume that the continuous slowing down approximation is valid, which requires that the rate of energy loss is the same as the total stopping power. In that case, the distance that a muon with initial kinetic energy $E_\mu$ travels before reaching the threshold energy $\Eth$, can be calculated as
\begin{equation}
R(E_\mu,\Eth) = \int_{E_\mu}^{\Eth} \frac{-dE}{\langle dE/dX \rangle}.   
\end{equation}
We assume the same energy threshold as in SuperK analyses, namely $\Eth\simeq1.6\GeV$.  The threshold energy corresponds to a muon of track length of 7~m in water, which is the minimum track length considered by SuperK in their upward-going muon analysis categories. We also define an upper threshold $E_{\rm{th}}^{\rm{max}}$. Muons which reach the detector with energies above $E_{\rm{th}}^{\rm{max}}$ have path lengths longer than the longest possible path through the inner detector (ID). For HyperK, this path length is 89.5~m and the corresponding energy is $E_{\rm{th}}^{\rm{max}}=22.0$~GeV.  For SuperK the energy is $E_{\rm{th}}^{\rm{max}}=11.7$~GeV.  In Fig.~\ref{fig:muonrange}, we show the muon range calculated using an interpolating function for the total average energy loss rate.

The  probability $P(E_\nu,\Eth)$ that a neutrino with energy $E_\nu$ produces a muon via a charged-current interaction, and that muon reaches the detector with an energy higher than the threshold $\Eth$, is 
\begin{equation}
P(E_\nu,\Eth)=  N_A \int_0^{1-\Eth/E_{\nu}} \int_0^1 \dfrac{d^2\sigma_\nu}{dx dy}  R(E_\mu,\Eth) dx \, dy,
\label{eq:prob}
\end{equation}
where $N_A$ is  Avogadro's number, $\sigma_\nu$ is the charged current neutrino-nucleon cross section,  $x$ and $y$ are the Bjorken scaling parameters, 
\begin{eqnarray}
x &=& \frac{Q^2}{2m_N E_\nu y},\\
y &=& 1 - \frac{E_\mu}{E_\nu},
\end{eqnarray}
$Q$ is the momentum transfer between neutrino and muon, and $m_N$ is the nucleon mass. 
As in the SuperK analyses, the total charged-current cross section is approximated in \texttt{GENIE}~\cite{Andreopoulos:2015wxa} by
\begin{equation}
\sigma_\nu^{\rm CC} \simeq \sigma_{\nu}^{\rm QEL} + \sigma_{\nu}^{\rm RES} +  \sigma_{\nu}^{DIS},
\end{equation}
where the three terms correspond to quasi-elastic scattering, baryon resonance production and deep inelastic scattering, respectively. We note that resonance production is dominated by the $\Delta$ resonance. 

The upward going muon flux can then be calculated as
\begin{equation}
\dfrac{d\Phi_\mu(\Eth,\cos z_\mu)}{d\Omega_\mu} = \int_{\Eth}^\infty dE_\nu \,  P(E_\nu,\Eth) \dfrac{d^2\Phi_\nu (E_\nu,\cos z_\nu) }{dE_\nu d\Omega_\nu}, 
\label{eq:totalupmuflux}
\end{equation}
where $z_\mu$ and $z_\nu$ are the zenith angles of the final muon and its parent neutrino, respectively. 
Assuming that the direction of the muon is approximately the same as that of its parent neutrino, we have
\begin{equation}
\dfrac{d\Phi_\mu(\Eth,\cos z)}{d\Omega} = \int_{\Eth}^\infty dE_\nu \,  P(E_\nu,\Eth) \dfrac{d^2\Phi_\nu (E_\nu,\cos z) }{dE_\nu d\Omega}. 
\label{eq:totalupmufluxapprox}
\end{equation}
We proceed in the same way with muon anti-neutrinos. 

\begin{figure}[t]
    \centering
\includegraphics[width=0.45\textwidth]{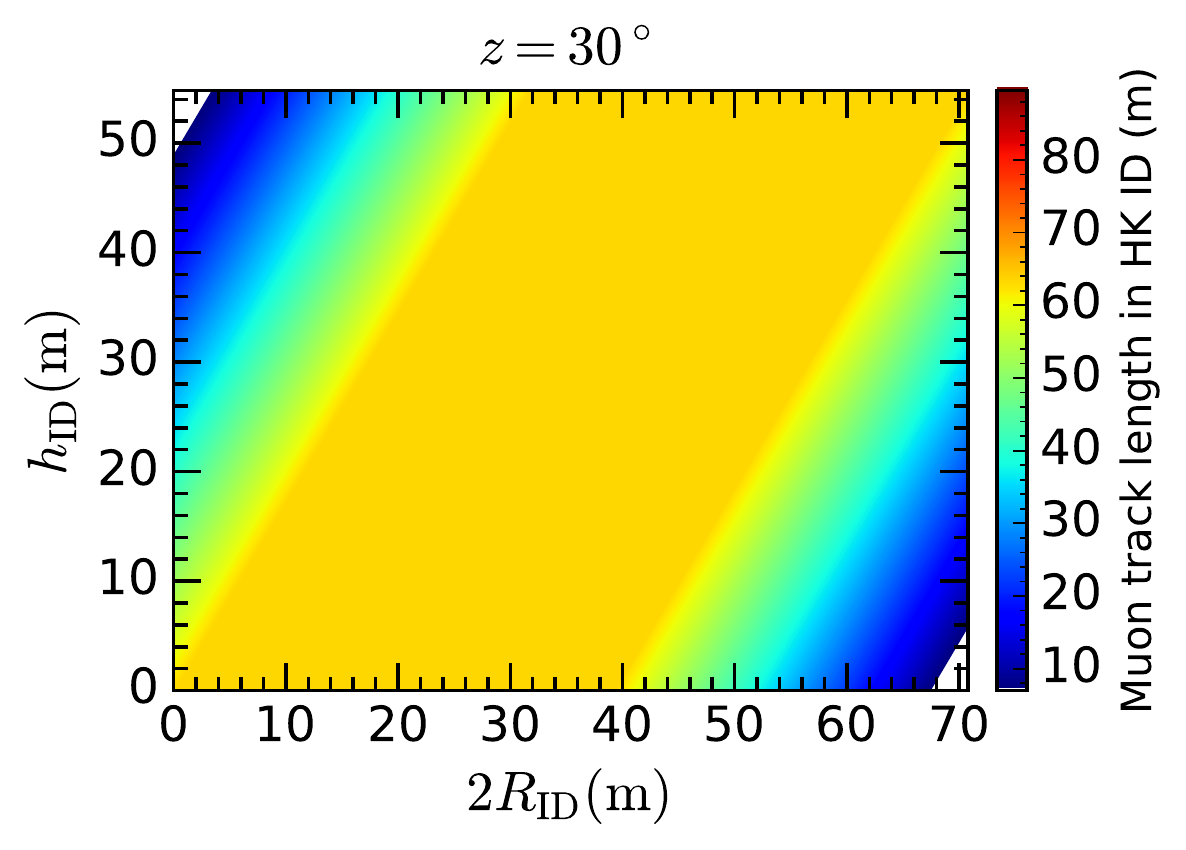}
\includegraphics[width=0.45\textwidth]{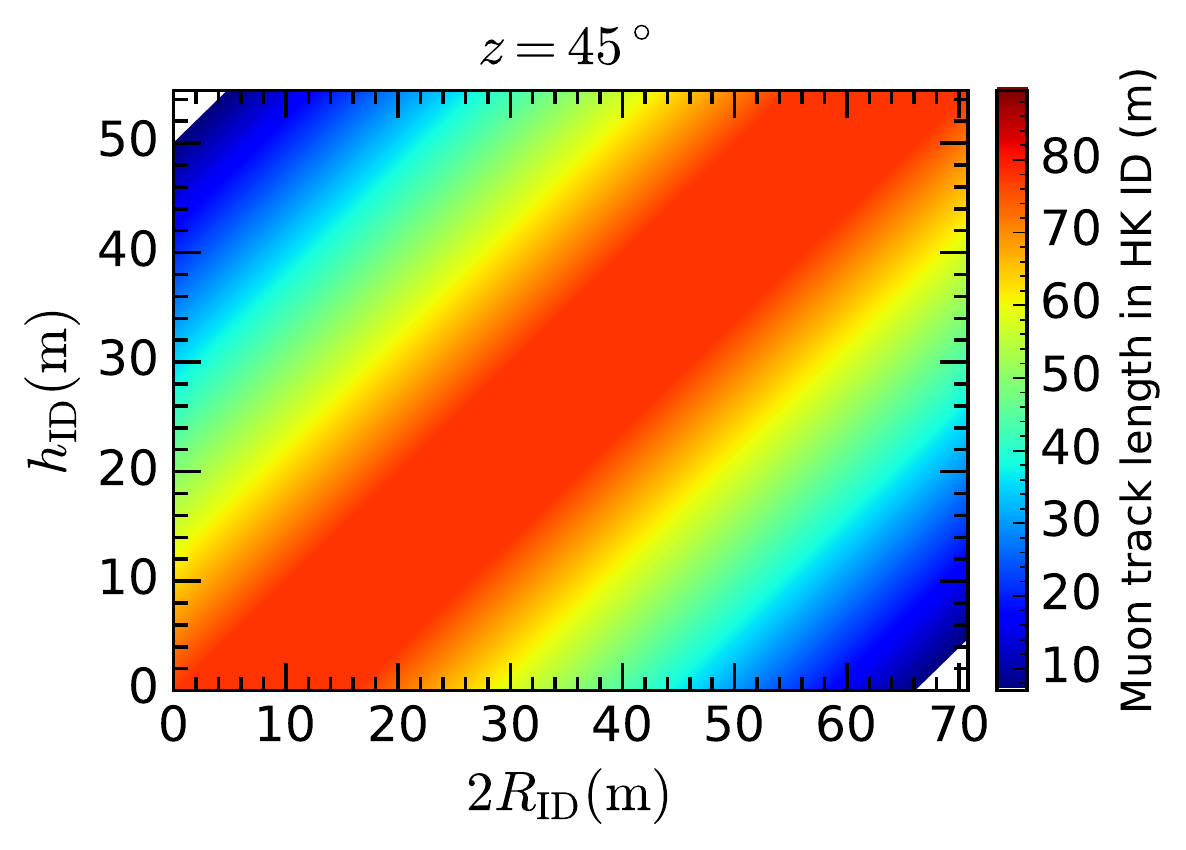}
\includegraphics[width=0.45\textwidth]{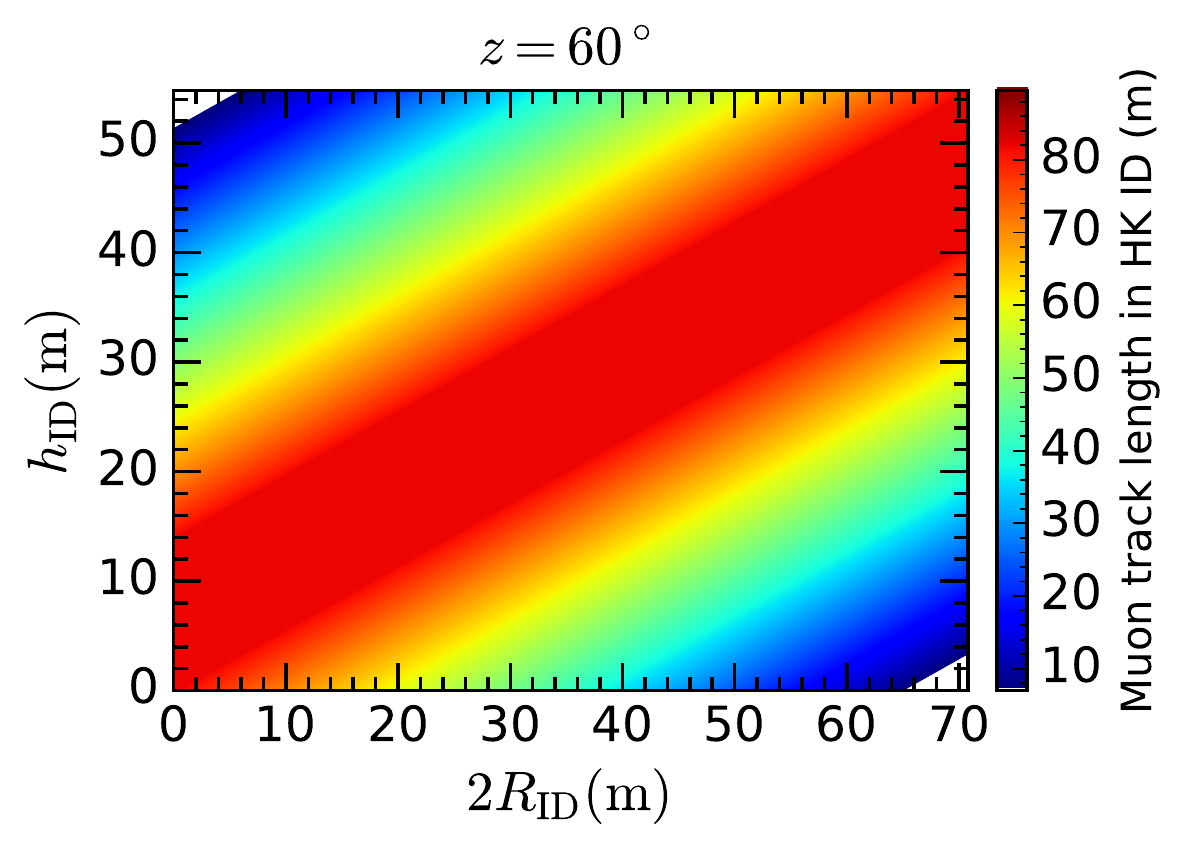}
    \caption{Maximum muon track lengths for upward through-going events in the HyperK  inner detector for different zenith angles ($z$=30, 45 and $60^\circ$). The  line spacing in the direction perpendicular to the zenith angle is $10\cm$ as described in the text. The minimum track length is 7m, and maximum track length is 89.5 m. }
    \label{fig:HKmaxmuontracks}
\end{figure}

The effective expected upward through going muon flux is calculated from Eq.~\ref{eq:totalupmufluxapprox} by averaging over all the possible muon path lengths  $x_i$, 
\begin{equation}
 \dfrac{d\Phi_\mu(\cos z)}{d\Omega} =  \dfrac{\sum_i \dfrac{d\Phi_\mu(\Eth(x_i),\cos z)}{d\Omega} \Theta(x_i-7\m)}{\sum_i \Theta(x_i-7\m)} , 
\end{equation}
where the maximum muon track length at  SuperK is 49.5~m, and at HyperK is 89.5~m.
Muon path lengths are computed at each point in a 2D grid  with $10\cm$ step-size (corresponding to the approximate detector resolution) in a plane perpendicular to the muon direction. This is equivalent to the detector effective area used to determine the observed flux.
In Fig.~\ref{fig:HKmaxmuontracks}, we show a transversal view of the HyperK inner detector, displaying the maximum track lengths ${\rm Max}[x_i(z)\geq7\m]$ for a given zenith angle. This is the same value as at SuperK, and could in principle be different at HyperK. We have varied the minimum track length between 6m and 8m and found it to have only a minor impact on our results. Note that every line in the figure corresponds to an ellipse or a truncated ellipse.  For very small values of $z$ the maximum track length approaches the inner detector diameter and for large values of $z$ it approaches the detector height.

\begin{figure}[t]
    \centering
\includegraphics[width=\textwidth]{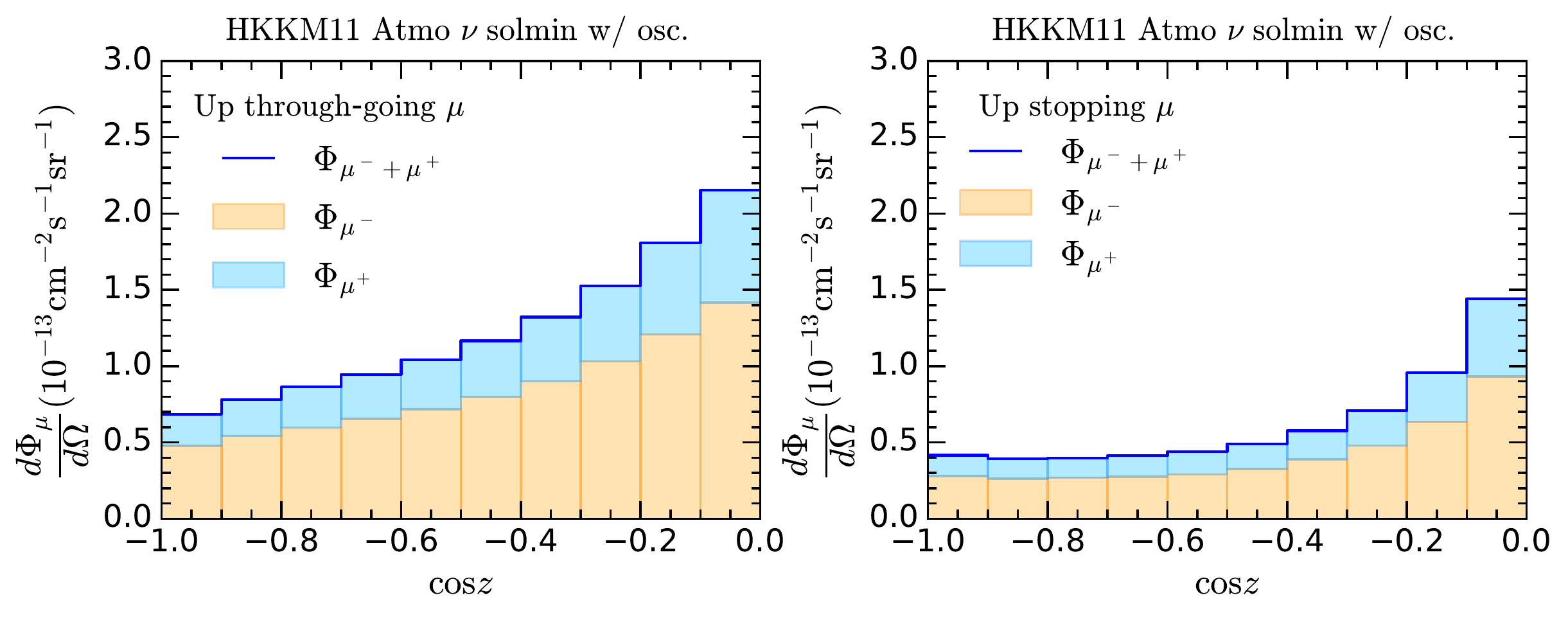}  
\caption{The expected upward through-going (left) and upward stopping (right) muon fluxes at HyperK, as a function of the zenith angle, for atmospheric neutrinos. Neutrino oscillation effects have been included. The individual contributions from $\mu^-$ and $\mu^+$ are shown in orange and light blue respectively, with the sum shown as the solid blue line.}
    \label{fig:upmu_atmofluxosc}    
\end{figure}

The upward stopping muon flux is obtained by subtracting the  through-going from the total upward-going muon flux, with the latter given by Eq.~\ref{eq:totalupmufluxapprox} with $\Eth\simeq1.6\GeV$. 
The atmospheric neutrinos fluxes at HyperK, obtained from the method above, are shown in Fig.~\ref{fig:upmu_atmofluxosc} for through-going muons (left) and stopping muons (right) as a function of $\cos z$. We show the contribution from muon and anti-muons separately, although water Cherenkov detectors such as HyperK are sensitive only to the sum, shown as the solid blue line. 

\begin{figure}[t]
    \centering
    \includegraphics[width=0.6\textwidth]{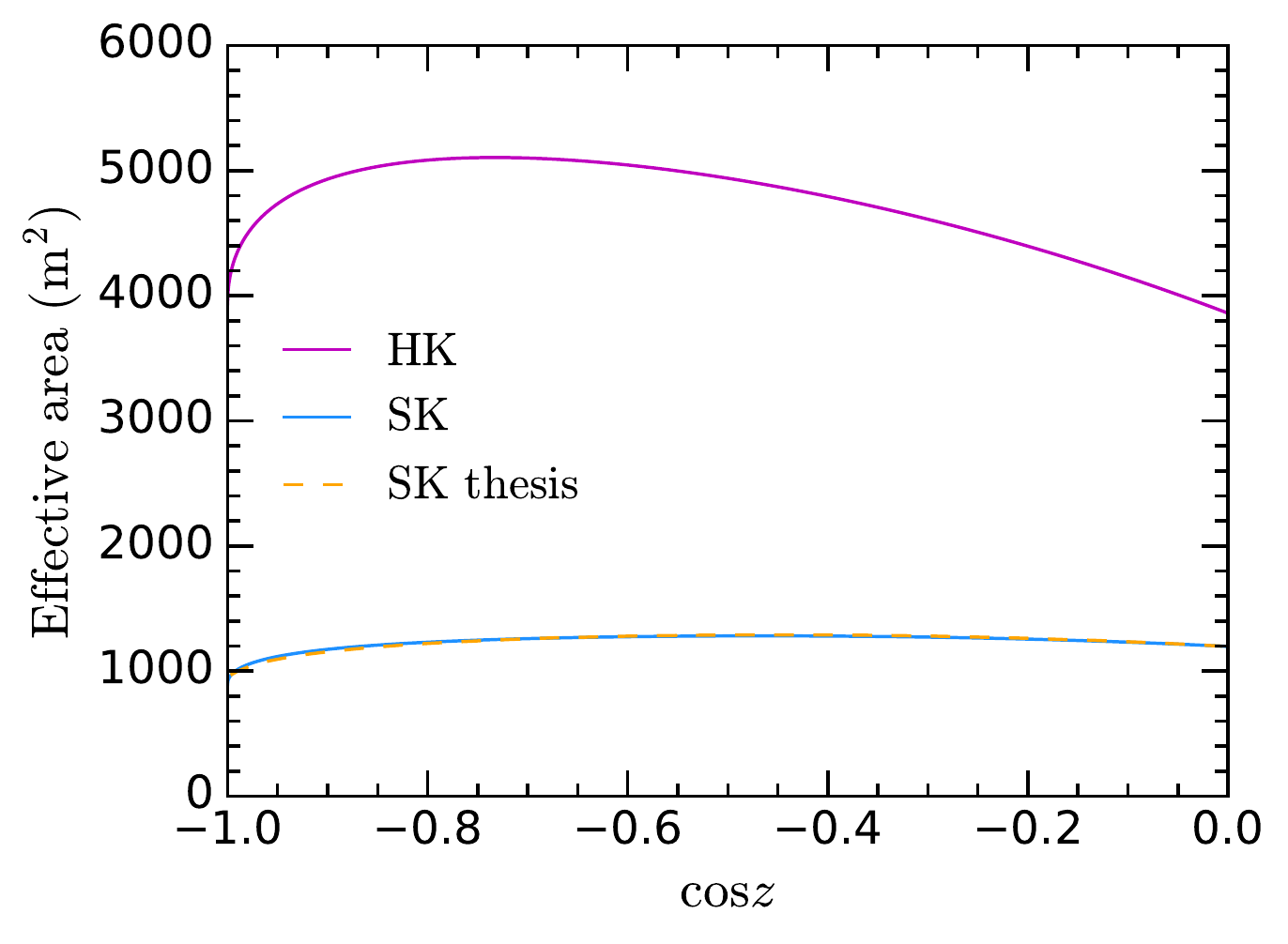}
    \caption{The effective area as a function of the zenith angle for upgoing muons with track lengths longer than $7\m$ for HyperK (top, magenta line) and SuperK (lower, blue line) derived using Eq.~\ref{eq:aeff}. Also shown is the effective area of SuperK taken from ref.~\cite{Tanaka:2011zz} (dashed orange), which is partially obscured by the blue line.}
    \label{fig:effectivearea}
\end{figure}

The fluxes in Fig.~\ref{fig:upmu_atmofluxosc} must be multiplied by the effective area of the detector, which varies with $\cos z$. The problem is then to calculate the cross-sectional area of a cylindrical detector as a function of the zenith angle, while taking into account the minimum 7~m track length. We denote the minimum track length by $l_{min}$, the height of the inner detector by $\hID$ and the radius of the inner detector by $\RID$. We define the angle $\theta(z, \RID)$ as
\begin{equation}
\theta(z,\RID) = \arccos \left( \frac{\RID-l_{min}\sin z}{\RID} \right) \,
\end{equation}
and an ancillary function as
\begin{equation}
A_{top}(z,\RID) = \pi \RID^2 \theta(z,\RID) - \left(\RID-l_{min}\sin z \right) \RID \sin\theta(z,\RID) \,.
\end{equation}
This corresponds to the area that must be subtracted from the area of the base due to the minimum path length $l_{min}$.
The effective area as a function of the zenith angle and parameters of the inner detector is then
\begin{multline}
    A_{\textrm{eff}}(z,\hID, \RID) = \left( \pi \RID^2 -2 A_{top}(z,\RID)\right)\cos z  \\ 
    + 2\sin z \left(\hID -l_{min} \cos z \right) \sqrt{\RID^2-\left(\frac{l_{min} \sin z }{2} \right)^2} \,.
    \label{eq:aeff}
\end{multline}
The effective areas for SuperK (blue solid line) and HyperK (magenta solid line) as functions of the zenith angle are shown in Fig.~\ref{fig:effectivearea}. For comparison, we also show the effective area of SuperK taken from Fig.~6.30 in the PhD thesis of ref.~\cite{Tanaka:2011zz} (orange dashed line), finding very good agreement. The effective are of HyperK is just over four times that of SuperK.

\begin{figure}
    \centering
    \includegraphics[width=0.55\textwidth]{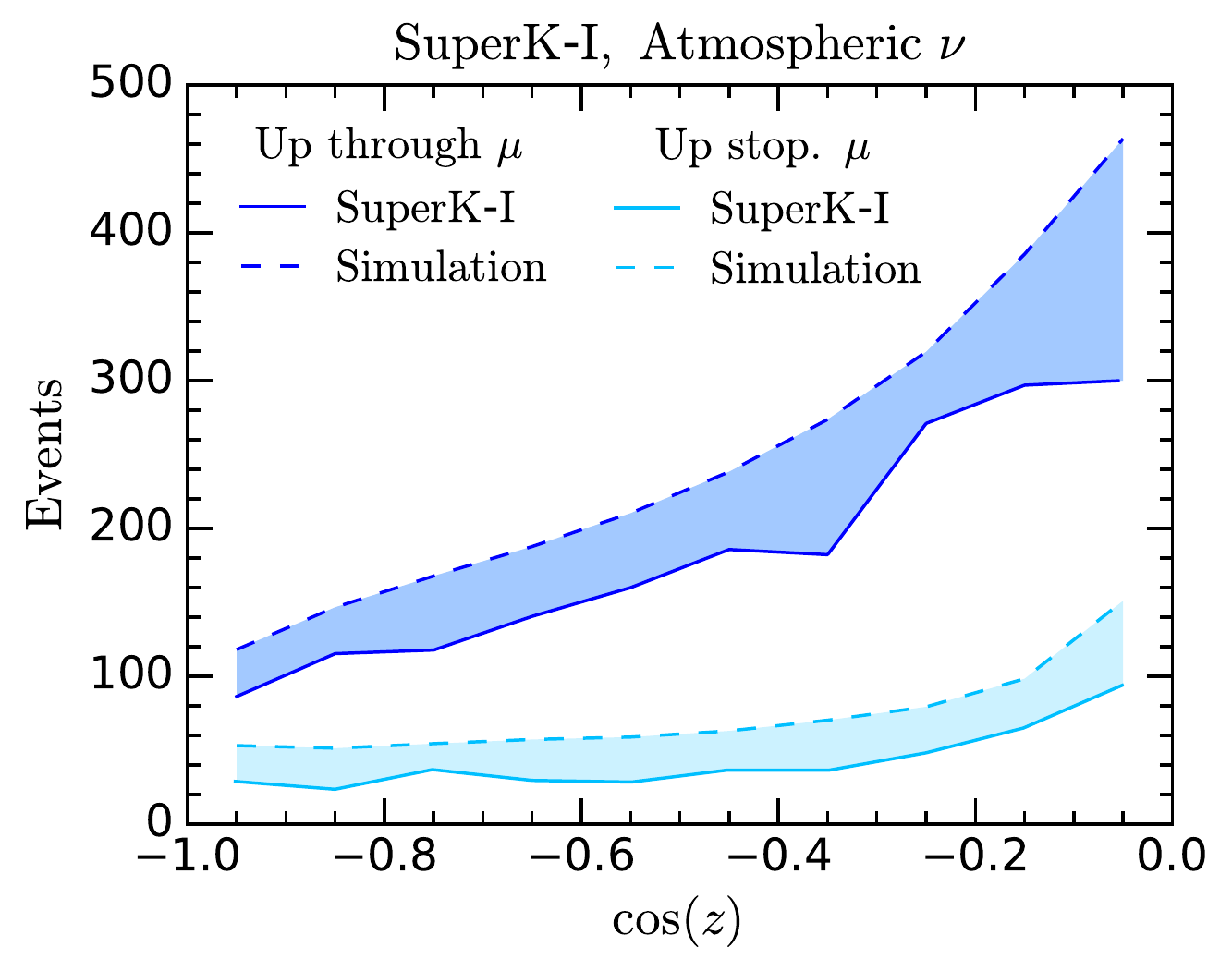}
    \caption{Comparison of our SuperK simulation (dashed) with SuperK-I atmospheric neutrino data~\cite{Tanaka:2011zz} (solid) for the upward-going muon categories. Light blue lines correspond to upward stopping muons, and the dark blue to upward through-going muons.}
    \label{fig:SK_comparison}
\end{figure}

To validate our workflow, we compare our results for atmospheric neutrinos with those in Fig.~6.31 in ref.~\cite{Tanaka:2011zz}, which shows  data from SuperK-I for stopping muons, non-showering through-going muons and showering through-going muons. We add the data from the two through-going muon categories together, since we do not discriminate between them. We multiply the fluxes of upward-going muons we calculated by the angle-dependent effective area, and the exposure time of 1646 live days. The results are shown in Fig.~\ref{fig:SK_comparison}. The upper, blue-dashed line is the number of through-going muon events from our simulation, while the upper solid line is the SuperK-I data. Similarly, the lower dashed and solid lines are our calculation and the SuperK-I results for stopping muons. In both cases our predictions are 30-40\% higher than the SuperK-I data, with the overall shape as a function of the zenith angle matching well. 

There are a number of differences in our calculation of the upward-going muon fluxes which may explain why they do not exactly match those from SuperK-I. Our calculation of the upward-going muon fluxes is semi-analytic rather than the large-scale Monte Carlo approach adopted by SuperK. There are also differences in the simulation software. We use \texttt{GENIE} v.3.0.4a while SuperK uses an older version of \texttt{NEUT}. We use the HKKM11 atmopheric neutrino fluxes, and SuperK use the older HKKM06~\cite{Honda:2006qj} fluxes. We use the same parton distribution functions (GRV98).

\begin{figure}
    \centering
\includegraphics[width=\textwidth]{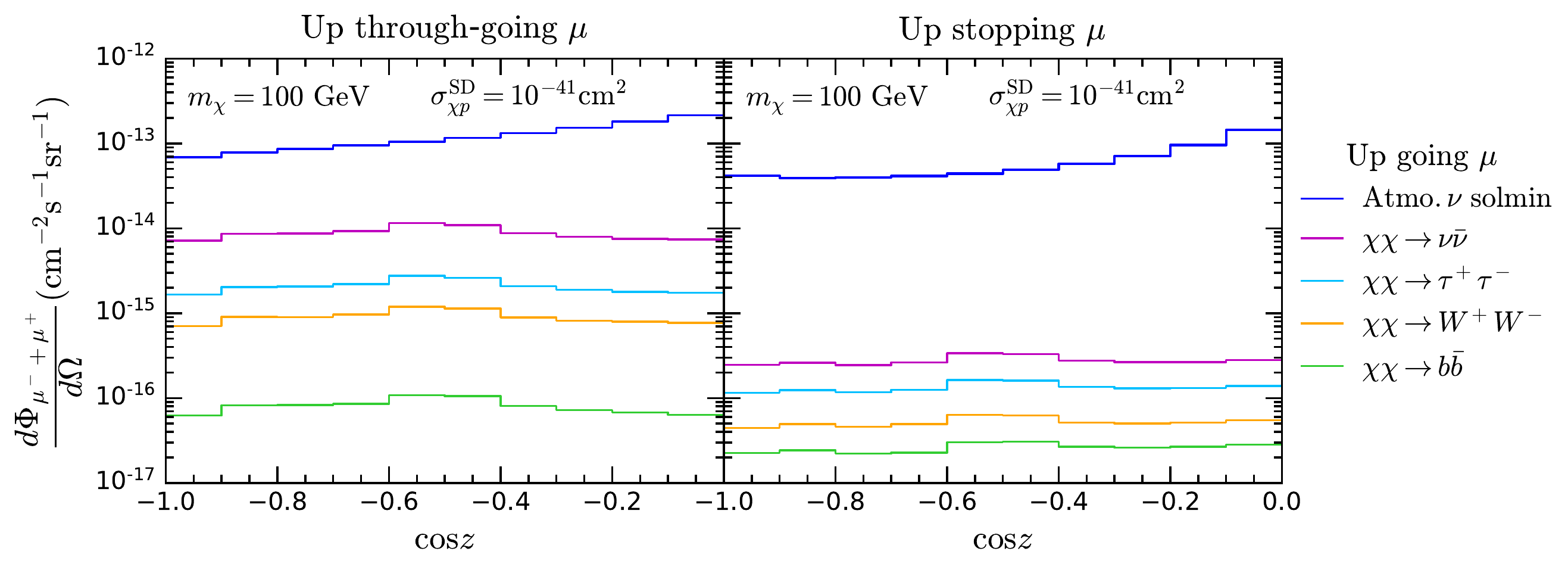}    
    \caption{Expected upward through going (top) and stopping (bottom) muon flux at HyperK for atmospheric neutrinos (blue) and solar DM annihilating to $\nu\overline{\nu}$ (magenta), $\tau^+\tau^-$ (light blue),  $W^+W^-$ (orange) and $b\Bar{b}$ (green) for $\mdm=100\GeV$ and $\sigmaSD=10^{-41}\cm^2$. }
    \label{fig:DMflux}
\end{figure}

Finally, to calculate the upward going muon flux from dark matter annihilation in the Sun, we first generate events in the rock surrounding the detector with \texttt{WimpEvent} and calculate the binned flux using Eq.~\ref{eq:DMfluxbinned}. We then use the same analytic method outlined above. The signal fluxes for dark matter annihilation into $b\Bar{b}$, $\tau^+ \tau^-$, $W^+ W^-$ and $\nu\overline{\nu}$ at HyperK are shown in Fig.~\ref{fig:DMflux} for through-going muons (left panel) and stopping muons (right panel) as a function of the zenith angle. For this figure, we have fixed the dark matter mass to $m_{\chi}=100$~GeV and the spin-dependent scattering cross-section to $\sigma^{SD}_{\chi p}=10^{-41}~\rm{cm}^2$. The largest neutrino flux is associated with DM annihilation directly to neutrinos, and the smallest with annihilation $b\Bar{b}$. As expected, the fluxes for through-going muons are larger than up-stopping muons for $m_\chi=100$~GeV.

\section{Results}
\label{sec:results}
 
The most recent SuperK search for dark matter annihilation in the Sun is ref.~\cite{Choi:2015ara}. This analysis includes nearly 11 years of live-data, accumulated during running of phases of SuperK-I through to SuperK-IV. That analysis utilised the Fully Contained (FC) data set for electrons and muons, Partially Contained (PC) $\nu_\mu + \bar{\nu}_\mu$, and the through-going and stopping-muons. We use our SuperK detector simulation from ref.~\cite{Bell:2020rkw} to reproduce the limits from ref.~\cite{Choi:2015ara}. For simplicity, we do not attempt to model the different acceptances of SuperK-I to IV, although these are not insignificant. As in ref.~\cite{Bell:2020rkw} we use the \texttt{Swordfish} statistics package~\cite{Edwards:2017kqw,Edwards:2017mnf} to derive 90\% confidence limits on the FC $\nu_e+\bar{\nu}_e$, FC $\nu_\mu+\bar{\nu}_\mu$, PC $\nu_\mu+\bar{\nu}_\mu$, through-going muon, and stopping upward-going muon event categories. We also derive a combined limit combining the information from all six categories. \texttt{Swordfish} calculates limits based on a maximum likelihood estimation following a novel reduction of the multi-bin problem to a single bin problem determined by an equivalent number of signal and background events. 

We use 20 bins in $E_{\rm{kin}}$ for the FC and PC categories, and 10 bins in $\cos(z)$ for the upward-going muon categories so that our combined analysis has 80 bins. The signal and background fluxes for the different categories are governed by the same normalisations, respectively. The uncertainties in each bin and each category will in general be different, and possibly correlated with one another. However, we assume that they are independent, and use a 30\% systematic uncertainty. SuperK has a substantially more sophisticated combination of categories and treatment of uncertainties. They use 18 subcategories and 480 bins, with between 48 and 66 different
sources of uncertainty (depending on whether the uncertainty is in the signal or background). For the parameter space we study, the combined limit is dominated by one event category (either the up-stop or up-through muons). We expect that the treatment of how the categories are combined should only matter in the regions where more than one event category makes an important contribution to the combined limit. This mostly occurs when the dominant category switches from up-stop to up-through at intermediate dark matter masses.  
 We use an exposure time of 4177.7 live days for all categories, as given in ref.~\cite{Choi:2015ara}. We note that the FC and PC categories have slightly fewer live days (3902.7) in the SuperK~I-IV analysis~\cite{Choi:2015ara}, but the difference is too small to affect our limits, which in any case are dominated by the two upward-going muon categories.

The SuperK Collaboration provide limits for a number of DM masses between 4~GeV and 200~GeV, for the $b\Bar{b}$, $\tau^+\tau^-$ and $W^+W^-$ final states. It is clear from their Figure 2 however that they have had a downwards fluctuation in their data at low masses. Accordingly, we take the  $1\sigma$ uncertainty band on their expected limit, and compare our results with that. SuperK does not provide a $2\sigma$ uncertainty band. 

\begin{figure}[t]
    \centering
\includegraphics[width=0.495\textwidth]{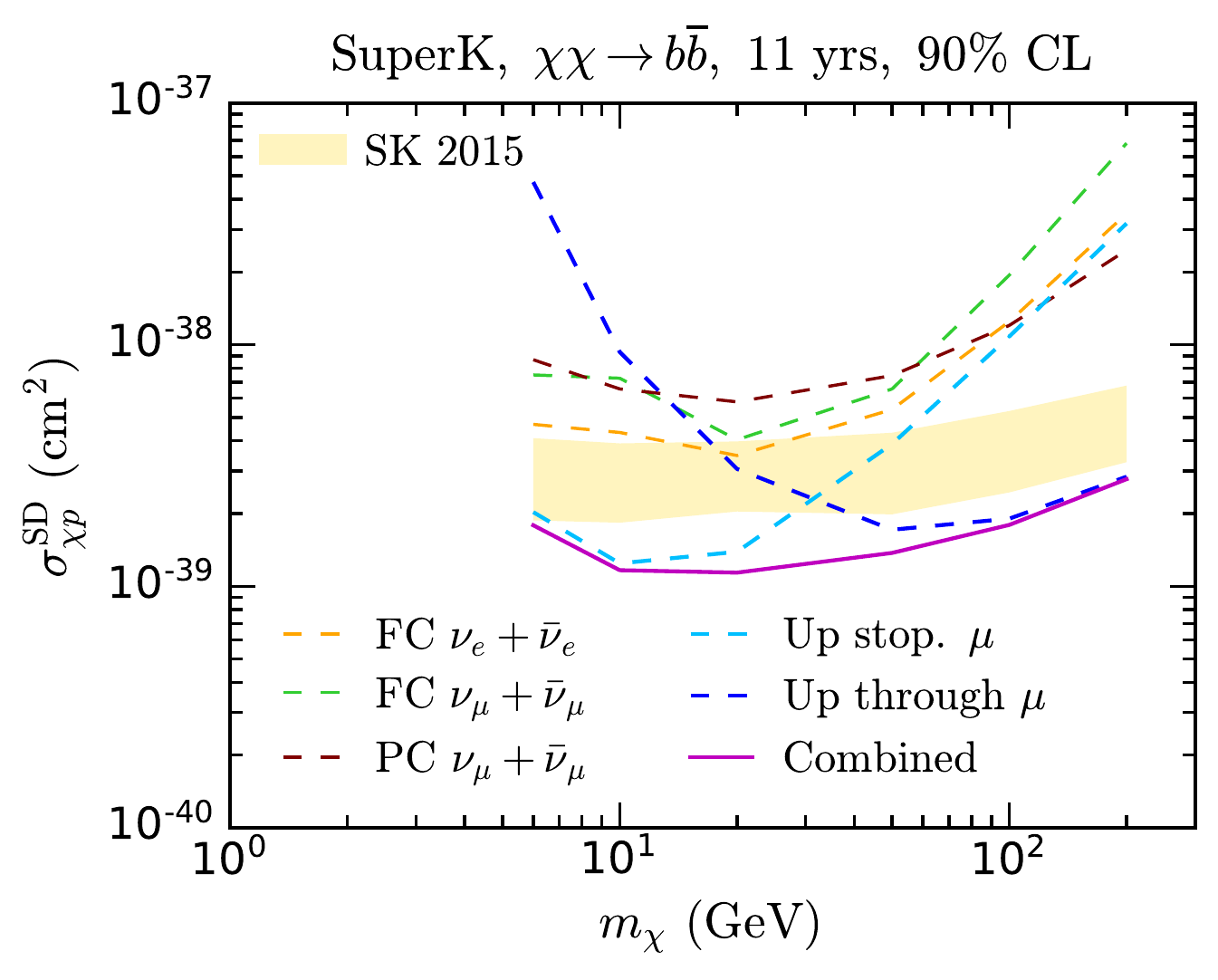}    
\includegraphics[width=0.495\textwidth]{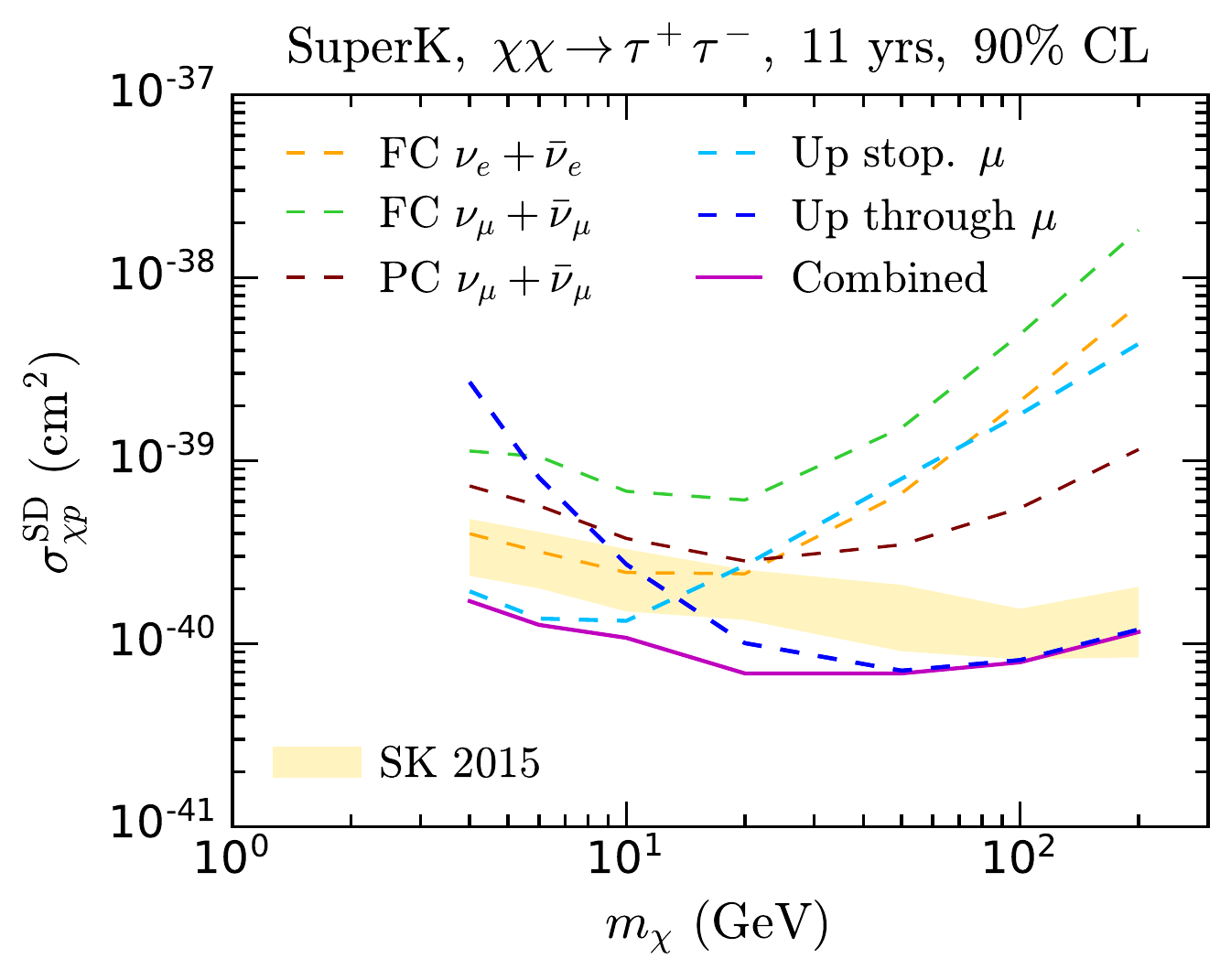}   
    \caption{Comparison between our projected limits and the expected limits from Super-Kamiokande for the $b\bar{b}$ (left) and $\tau^+\tau^-$ (right) final states. The SuperK $1\sigma$ expected limit is the yellow shaded region and our combined limit is the solid magenta line. Limits for individual categories are: FC $\nu_e+\bar{\nu}_e$ (orange dashed), FC $\nu_\mu+\bar{\nu}_\mu$ (green dashed), PC $\nu_\mu+\bar{\nu}_\mu$ (brown dashed), upward stopping muons (light blue, dashed), and upward through-going muons (dark blue dashed).}
    \label{fig:SKbb_tautau_compare}
\end{figure}

\begin{figure}[t]
    \centering
\includegraphics[width=0.495\textwidth]{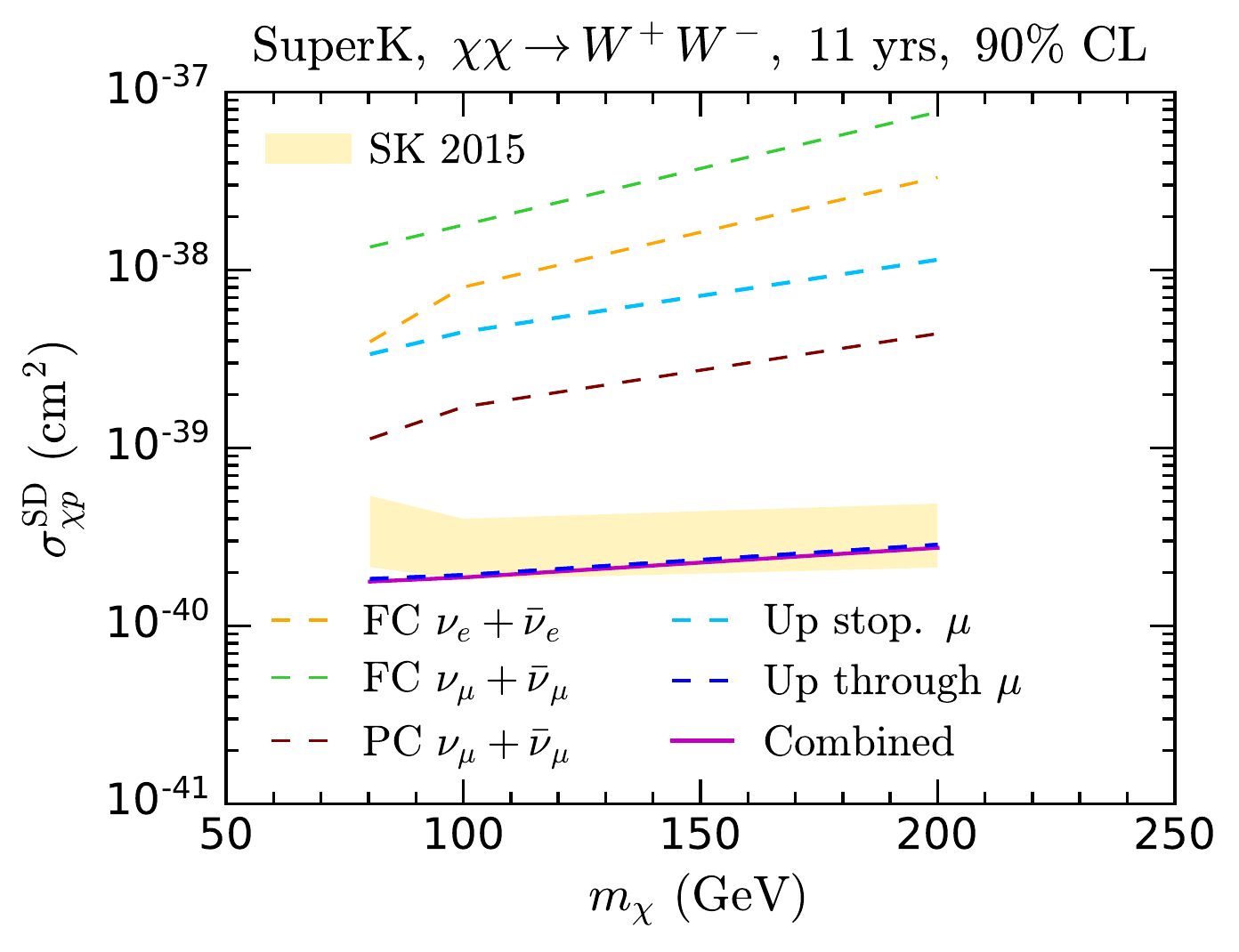}    
    \caption{Comparison between our projected limits and the expected limits from Super-Kamiokande for the $W^+ W^-$ final state. The SuperK $1\sigma$ expected limit is the shaded yellow region and our combined limit is the solid magenta line. Limits for individual categories are: FC $\nu_e+\bar{\nu}_e$ (orange dashed), FC $\nu_\mu+\bar{\nu}_\mu$ (green dashed), PC $\nu_\mu+\bar{\nu}_\mu$ (brown dashed), upward stopping muons (light blue, dashed), and upward through-going muons (dark blue dashed).}
    \label{fig:SKWW_compare}
\end{figure}

We show the SuperK expected limits as shaded yellow regions in Fig.~\ref{fig:SKbb_tautau_compare} ($b\Bar{b}$ on the left, $\tau^+\tau^-$ on the right)  and in Fig.~\ref{fig:SKWW_compare} (for $W^+W^-$). We also show the limits we obtain using the individual event categories: FC $\nu_e+\bar{\nu}_e$ (orange dashed), FC $\nu_\mu+\bar{\nu}_\mu$ (green dashed), PC $\nu_\mu+\bar{\nu}_\mu$ (brown dashed), upward stopping muons (light blue dashed), and upward through-going muons (dark blue, dashed).


For low DM masses, the most important event category is the stopping muons. As the dark matter mass increases, so does the resulting neutrino energy and hence the muon energy. At higher DM masses, above a few tens of GeV, the through-going muons become more important (see also Fig.~\ref{fig:muonrange}). Since the $W^+W^-$ channel is kinematically open only for DM masses greater than $m_W$, the only relevant events in that case are from through-going muons. 

Our combined limits are shown as solid magenta lines, to be compared with the SuperK limits shown by shaded yellow regions. While our limits generally reproduce the shapes of those from SuperK, we note that they are consistently somewhat stronger than those from SuperK. However, our results intersect the $1\sigma$ band in all cases, and given the size of the $1\sigma$ band, would probably be within the $2\sigma$ band. The SuperK Collaboration does not use the same limit-setting procedure we do, and has a far more sophisticated treatment of uncertainties (described in some detail in ref.~\cite{Abe:2014gda}). Along with their different treatment of the backgrounds described above, we use a different software version to generate the signal. Specifically, we use \texttt{WimpSim} 5.0 and they use \texttt{WimpSim} 3.01.

\begin{figure}
    \centering
\includegraphics[width=\textwidth]{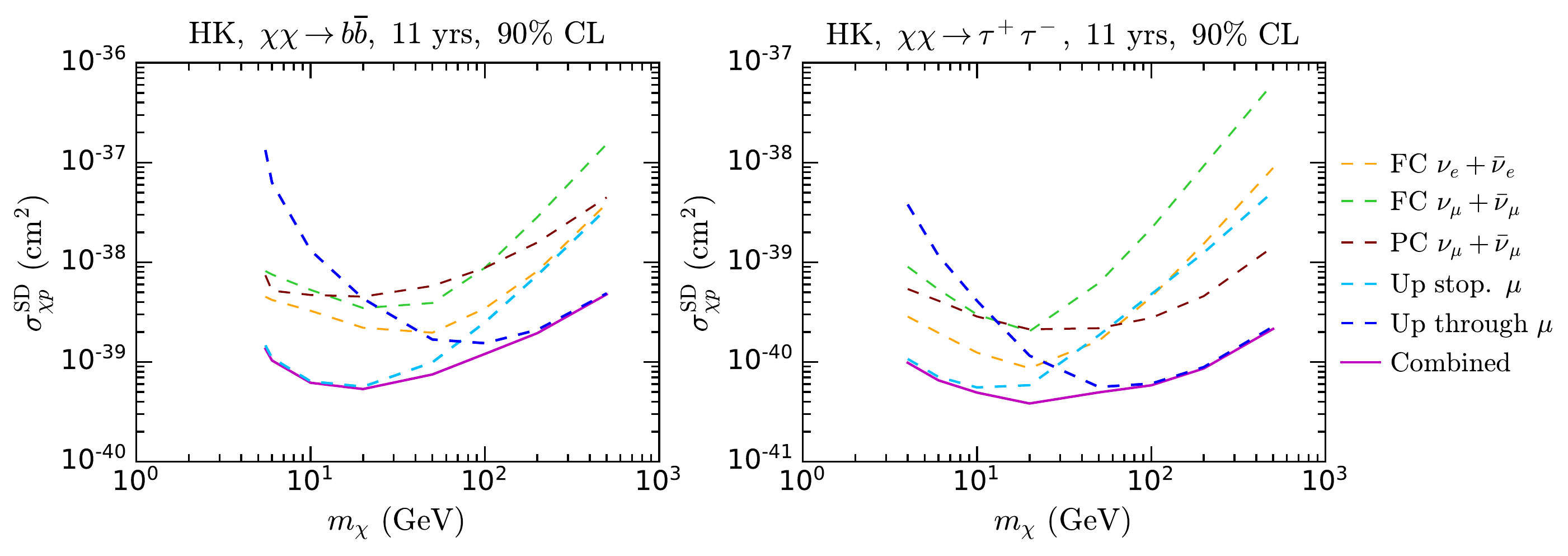}    
\includegraphics[width=\textwidth]{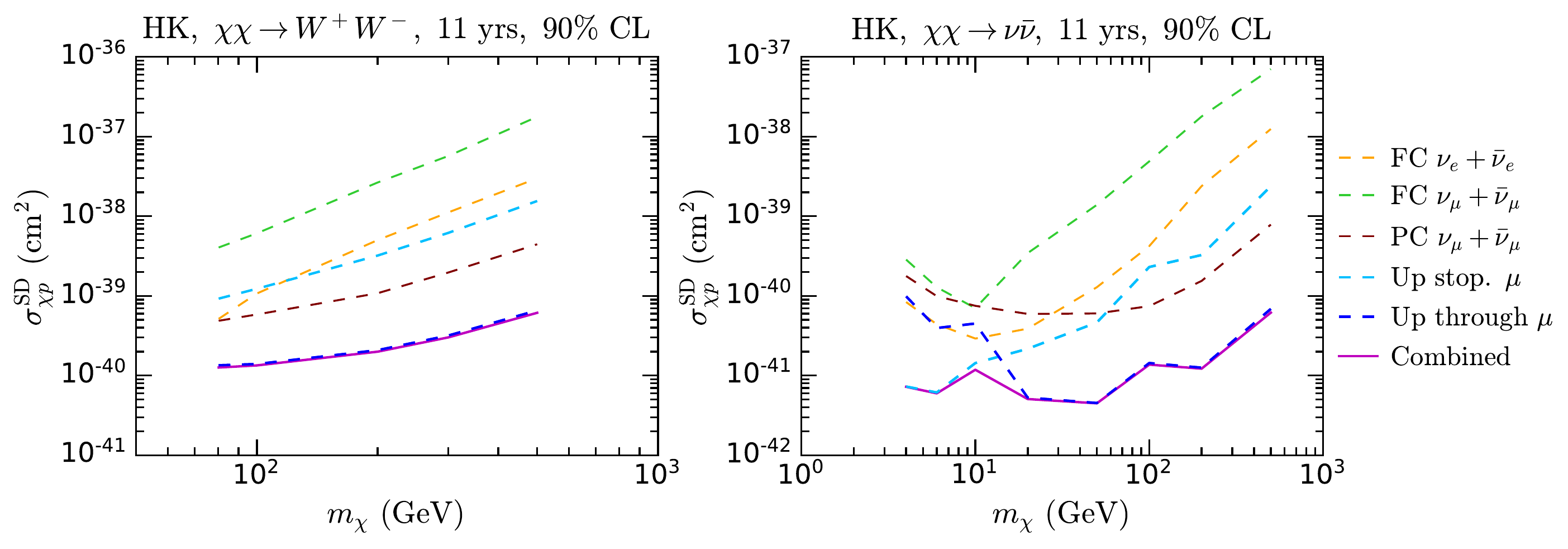}    
    \caption{Projected limits for the $b\bar{b}$, $\tau^+\tau^-$, $\nu\bar\nu$ and $W^+W^-$ (clockwise from top-left) final states at Hyper-Kamiokande. We show our combined HyperK projections (solid purple lines) and the projected limits for individual categories: FC $\nu_e+\bar{\nu}_e$ (orange dashed), FC $\nu_\mu+\bar{\nu}_\mu$ (green dashed), PC $\nu_\mu+\bar{\nu}_\mu$ (brown dashed), upward stopping muons (light blue, dashed), and upward through-going muons (dark blue dashed).}
    \label{fig:HK_all}
\end{figure}

In Fig.~\ref{fig:HK_all} we show our projections for the Hyper-Kamiokande experiment as magenta solid lines. The colour-coding of the lines is the same as in Fig.~\ref{fig:SKbb_tautau_compare}.  We show projections up to $m_{\chi}=500$~GeV. Above this mass range, other experiments such as IceCube are expected to have greater sensitivity. We also provide projections for the $\chi\chi\to\nu\bar{\nu}$ annihilation channel (bottom-right panel), which was not studied by SuperK. 

Our HyperK projections assume an exposure time of 4177.7 days (around 11 years, the same as in ref.~\cite{Choi:2015ara}) and the same systematics as for our SuperK simulation. We find that HyperK will improve on the current SuperK bounds by a factor of 2-3 depending on the dark matter mass. For comparison, the HyperK Design Report finds that the search for neutrinos from dark matter annihilating in the Earth will improve by a factor of 3-4 relative to SuperK (assuming no improvement in systematics over SuperK, and a 10 year exposure). Note that HyperK may run for longer than 11 years. An improved understanding of the systematics would also lead to stronger bounds on the properties of dark matter

\begin{figure}[t]
    \centering
\includegraphics[width=\textwidth]{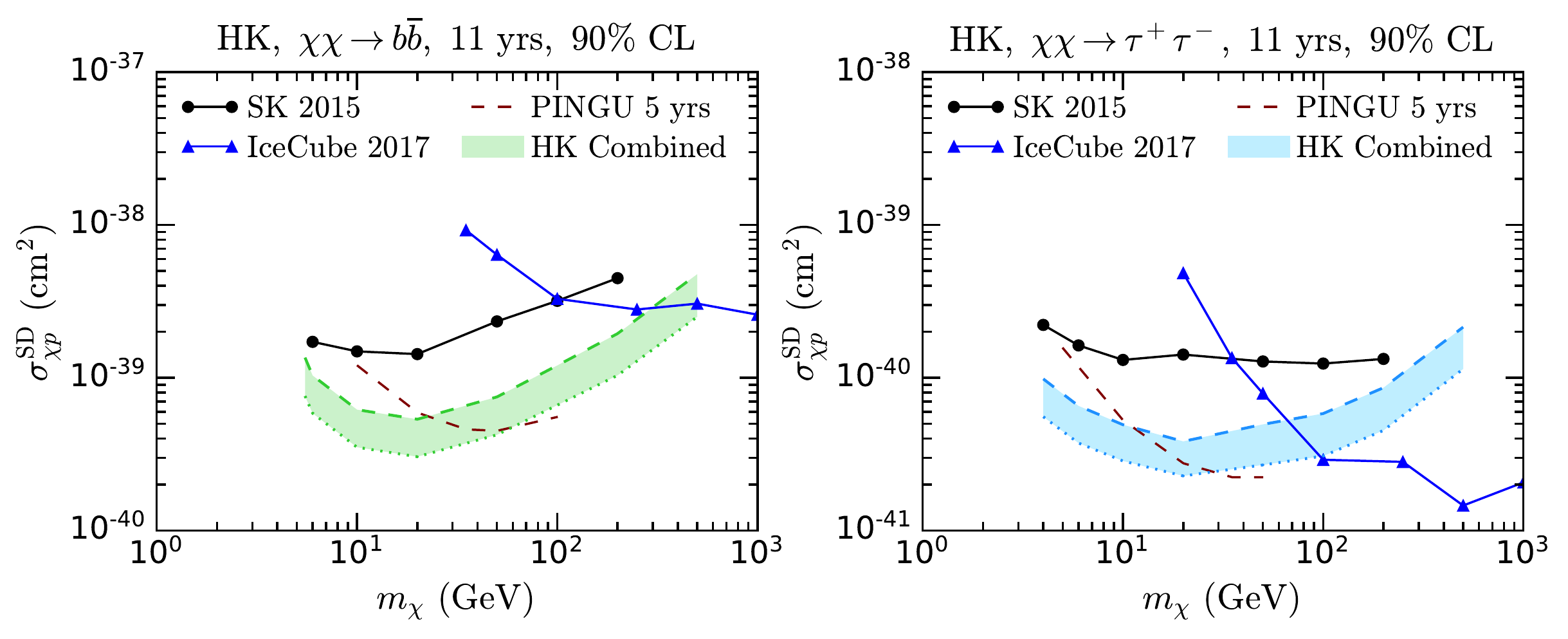}    
\includegraphics[width=\textwidth]{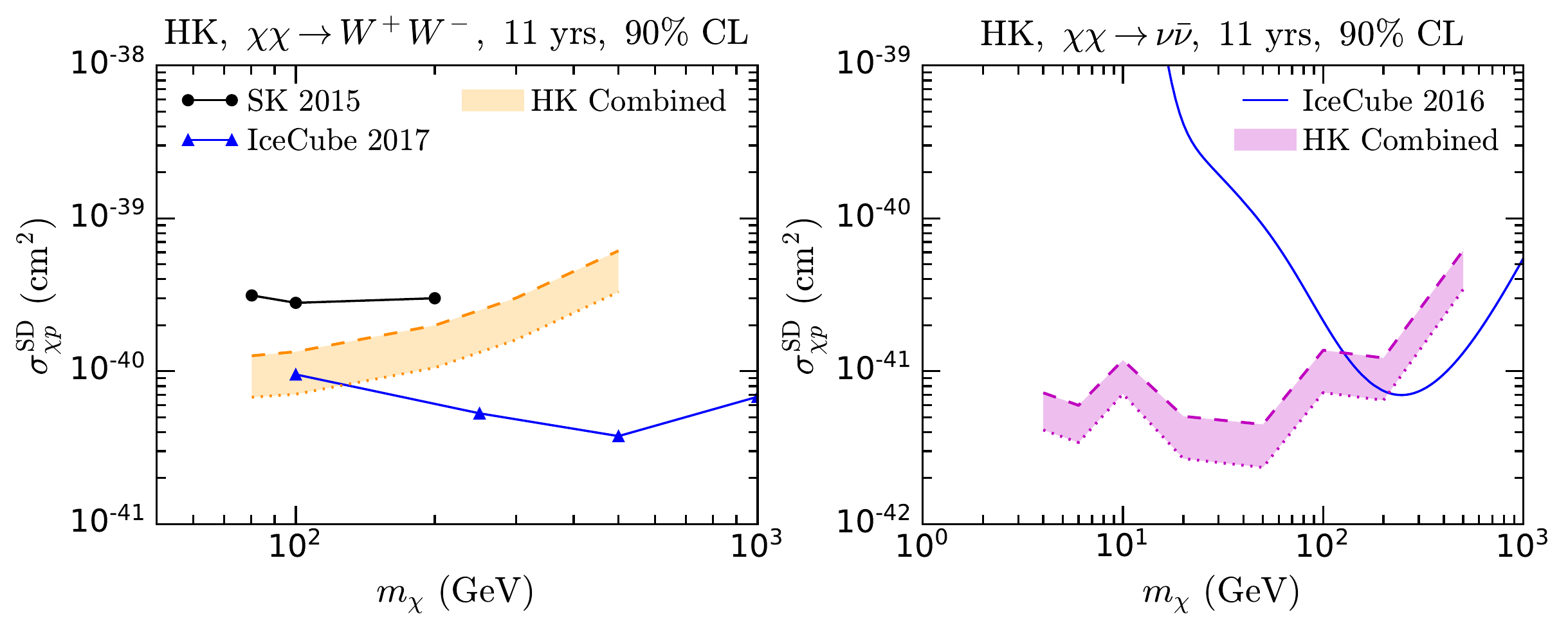}   
    \caption{Comparison of our HyperK combined projections with the same systematics used in Fig.~\ref{fig:SK_comparison} (dashed lines) with those for half the systematics (dotted lines). We also show the SuperK results (black solid, circles), limits from IceCube~\cite{Aartsen:2016exj,Aartsen:2016zhm} (blue solid, triangles), and projections for 5 years of PINGU data~\cite{Aartsen:2017mnf} (brown dashed).}
    \label{fig:HK_all2}
\end{figure}

We illustrate this point in Fig.~\ref{fig:HK_all2}, which compares our projections for the HyperK limits using the same systematics as we used for SuperK in Fig.~\ref{fig:SK_comparison} (the upper dashed lines), with those for the case where the systematics are halved (the lower dotted lines). For comparison, we also show limits for these final states from the IceCube experiment~\cite{Aartsen:2016zhm} (solid blue lines with triangles), and from ref.~\cite{Aartsen:2016exj} for $\nu\bar{\nu}$. While HyperK will probe new parameter space for annihilation to $b\bar{b}$, $\tau^+\tau^-$ and $\nu\bar{\nu}$ final states, most of the $W^+W^-$ parameter space accessible at HyperK has already been excluded by IceCube. In addition, IceCube already constrains the accessible parameter space above 100~GeV for $\tau^+\tau^-$ and $\nu\bar{\nu}$. We also note that our projections for HyperK correspond to the mid-2030s, by which time new data from IceCube will exist, and a next-generation experiment (PINGU or IceCube-Gen2) may be operational~\cite{TheIceCube-Gen2:2016cap,Aartsen:2020fgd}.  We therefore show projected limits for five years of PINGU operation for $b\bar{b}$ and $\tau^+\tau^-$ final state, taken from ref.~\cite{Aartsen:2017mnf}. The PINGU sensitivity will be similar to HyperK for masses above a few tens of GeV. Note that projections for the sensitivity of 3 years of PINGU operation, expressed in terms of the effective theory of DM-nucleon interactions, can be found in ref.~\cite{Backstrom:2018plo}.

\begin{figure}
    \centering
    \includegraphics[width=0.8\textwidth]{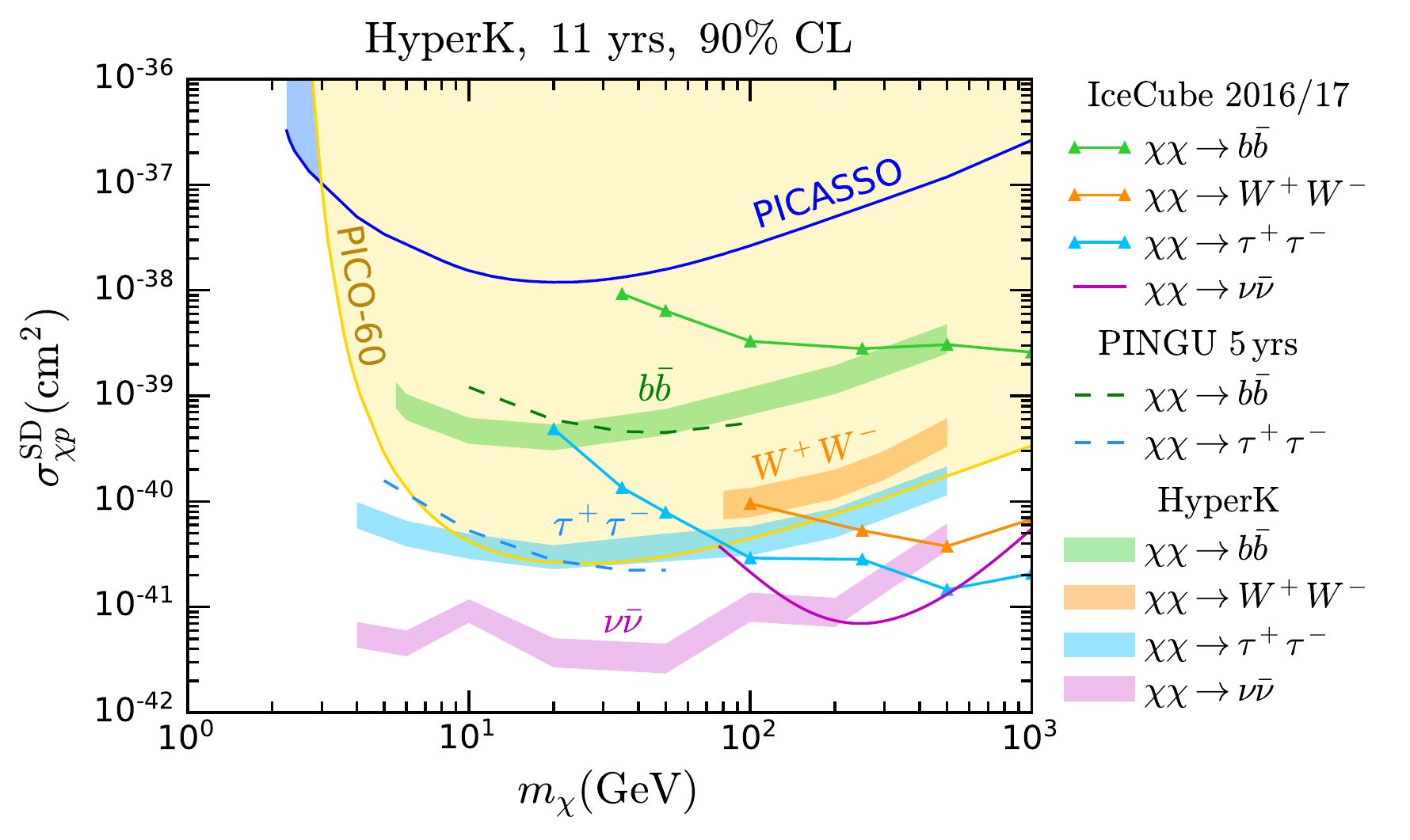}  
    \caption{Comparison of our HyperK projections with the direct detection limits from PICASSO~\cite{Behnke:2016lsk} (blue line) and PICO-60~\cite{Amole:2019fdf} (yellow line and shaded region). Also shown are the limits from IceCube~\cite{Aartsen:2016exj,Aartsen:2016zhm} and the projections for PINGU~\cite{Aartsen:2017mnf}.}
    \label{fig:compare_all}
\end{figure}

We present an alternate version of our results in Fig.~\ref{fig:compare_all}, including the leading spin-dependent proton scattering  cross-section constraints from direct detection experiments, namely PICASSO~\cite{Behnke:2016lsk} (solid blue line, relevant at low masses) and PICO-60~\cite{Amole:2019fdf} (yellow line and shaded region). We see that the relevant regions of parameter space for the $b\bar{b}$ and $W^+W^-$ final states are already ruled out by PICO-60. The HyperK reach for $\tau^+\tau^-$ is similar to the current PICO-60 limits, and the $\nu\bar{\nu}$ final state is substantially below the current constraints. We also include the IceCube results and PINGU projections discussed in the previous paragraph. While the direct detection limits rapidly weaken for DM masses below about 10~GeV,
evaporation of DM from the Sun becomes significant for DM masses below about 4~GeV, strongly suppressing any annihilation signal.

\section{Summary}
\label{sec:summary}

We have studied the prospects for the Hyper-Kamiokande experiment to measure a neutrino signal produced by the annihilation of dark matter captured in the Sun. We extended our simulations of the SuperK and HyperK detectors~\cite{Bell:2020rkw}, by implementing the upward-going stopping and through-going muon event categories, finding good agreement in validation against SuperK-I atmospheric neutrino measurements. We then simulated neutrino signals for DM annihilation in the Sun, to determine the HyperK sensitivity to the 
dark matter spin-dependent scattering cross section.

We applied our method to project the sensitivity for 11 years of exposure time at the HyperK detector (the same exposure as the SuperK search of ref.~\cite{Choi:2015ara}). We find that HyperK will be able to set limits which improve on the published SuperK results by a factor of two to three. This is a slightly smaller improvement than the factor of three to four estimated in the HyperK Design Report~\cite{Abe:2018uyc} for searches based on dark matter annihilation within the Earth. 
As in the projections of ref.~\cite{Abe:2018uyc}, we have assumed identical systematics for HyperK and SuperK. A reduction in these systematic uncertainties would lead to better limits. However, much of the parameter space accessible at HyperK is already constrained by direct detection experiments such as PICO-60~\cite{Amole:2019fdf}, or by IceCube~\cite{Aartsen:2016exj,Aartsen:2016zhm}.

\section*{Acknowledgements}
NFB, MJD and SR were supported by the Australian Research Council through the ARC Centre of Excellence for Dark Matter Particle Physics, CE200100008.

\bibliographystyle{JHEP} 
\bibliography{references}

\end{document}